# Classicity from Entangled Ensemble States of Knotted Spin Networks. A Conceptual Approach.[1]

**Rainer E. Zimmermann**

IAG Philosophische Grundlagenprobleme, FB 1, UGH,
Nora-Platiel-Str.1, D - 34127 Kassel.
Clare Hall, UK - Cambridge CB 3 9AL.[2]
Lehrgebiet Philosophie, FB 13 AW, FH,
Lothstr.34, D - 80335 München.[3]
pd00108@mail.lrz-muenchen.de

July 12, 2000

## Abstract

Referring to a conception put forward by Stuart Kauffman in his „Investigations", it is shown how the onset of classicity can be visualized in terms of an emergent process originating in entangled ensemble states of knotted spin networks. The latter exhibit a suitable autocatalytic behaviour effectively producing knots by knots acting upon other knots. In particular, a quantum computational structure can be described underlying spin networks such that most conditions for a partial ordering are not more valid for the latter. A conceptual argument is given then indicating that on a fundamental level, physics is non-local and a-temporal, and hence does not admit of the concept of causality. Hence, modelling the emergence of classicity in terms of a percolating web of coherence eventually decohering (in using a cellular automata architecture) does not imply the necessity of visualizing histories of directed percolation as causal sets. These aspects are compatible though with the recent concept of spin foams which do not actually imply distinguished directions of time flow on the micro-level.

---

[1] Contribution to icmp2000, 17-22 July 2000, Imperial College, London (GRL 2.11).
[2] Permanent addresses.
[3] Present address.

# 1 Introduction

In his 1996 paper on „Investigations" Stuart Kauffman [20] sets out to explore the possibility of general laws governing the class of coevolutionary self-constructing communities of autonomous agents. What he advocates in particular is a „fourth law" of thermodynamics stating that the evolutionary rule completing the other thermodynamical laws is showing up as a tendency describing the flow from the already actualized states to the adjacent possible states of some system, in the sense that the latter's dimensionality, on average, expands as rapidly as it can, thus leading to some kind of optimal fitness critically self-organized as if by an invisible hand. In fact, the idea is to think of this expansion as constituting what we generally call „time's arrow".

What is interesting for us here is that Kauffman tries to link these general ideas about the biosphere as we know it to the Universe altogether basing his approach mainly on two recent aspects of the research being undertaken on the fundamental structure of physics, namely on aspects of spin networks in particular. One aspect is Lee Smolin's idea of visualizing, on that fundamental level, quantum gravity as being definable in terms of spin networks which form combinatorially complex „knotted" structures. Hence, classical space (and mass-energy) can be thought of as crystallization and propagation of „seeds of classicity" as represented by Planck scale agents of the above mentioned type. See also [21] for more recent details.

On the other hand, referring to Lee Smolin's book „The Life of the Cosmos" [33], Kauffman argues in favour of cosmological selection, in the sense that the Universe is self-tuning its physical parameters such that it will maximaize its complexity. The idea is then that the expected maximal growth of the Adjacent Possible in the flow of a non-ergodic Universe maximizes the rate of quantum decoherence and thus the emergence of classicity.

In this present paper though, we will not discuss cosmological selection which has been done elsewhere [41]. In fact, recently, an indication has been uncovered which points towards this principle's plausibility, if however in a somewhat modified form [49]. Also, we will not discuss the implications of this model with respect to the Bohm-Hiley approach to quantum mechanics, which is a problem in its own right and will be discussed later at another place [44], [7].

What we will do is basically the following: First of all, we will recall some details about spin networks and spin foams and try to clarify some consequences in contextual terms (section 2). We will discuss then how entangled ensemble states of knotted spin networks could be actually visualized. And we will give a description of the relationship between the concept of quantum computation and these network structures (section 3). And finally, we will discuss some implications of this as to the appropriateness, adequacy, and feasibility of space and time as framework categories (including derived categories such as causality) on the fundamental level of physics (section 4). Note that we are essentially dealing with conceptual arguments only rather than giving technical details as to innovative insight into mathematical physics proper, in order to demonstrate what it actually is that philosophy can offer to physics nowadays and to show the stringency of philosophical conceptualization as a powerful heuristic instrument in its own right. Hence, this present paper is following a line of argument which has been started earlier when motivating the need of a modern philosophy of physics being visualized itself within a classical tradition of thinking [42], [43].

## 2  Spin Networks

Remember that spin networks are essentially trivalent graphs, i.e. graphs with three-valued nodes and spin values on the edges, their states being denoted by $|\Gamma>$. To take the norm refer to the mirror image of a graph and tie up the ends, forming a closed spin network $\Gamma\#\Gamma^*$ of value V such that

$$<\Gamma|\Gamma> = V(\Gamma\#\Gamma^*)$$

with

$$V(...) = \Pi\ (1/j!)\ \Sigma\ \epsilon(-2)^N,$$

here j being the edge label, N the number of closed loops, and $\epsilon$ referring to the intertwining operation taking care of permutations of signs. The product is to be taken over all edges, the sum over all routings. Hence, the networks can be visualized in diagrammatic form such as to represent the underlying „spin dynamics". For instance, a diagram with vertices a, b, c, and i, j, k within the region of intertwiners such that i + j = a, i + k = b, j + k = c, can be interpreted as two particles with spin a and b which produce (create) a particle with spin c. Spin interactions of this kind can lead to the creation of new structures: Take a large part of the network and detach from this small m-units, and n-units, respectively, as „free ends". The outcome of their tying up to form a new structure can be estimated in terms of a probability for the latter having spin number P, say. This turns out as being basically the quotient of the norm of the closed network and the norm of the network with free ends (times some intertwining operations). The *spin geometry theorem* tells us then that when repeating this procedure and getting the same outcome, then the new quotient is proportional to (1/2) cos θ, where the angle is one which is taken between the axes of the large units. Hence, it is possible to show that angles obtained in this way satisfy the well-known laws of Euklidean geometry. In other words: This purely combinatorial procedure can be used to actually approximate space from a *pre-spatial* structure which is more fundamental. This idea has been introduced by Penrose who originally tried to base his concept of *twistors* on spin networks. A generalization of spin networks and a connection with knot theory have been achieved more recently by Carlo Rovelli and Lee Smolin referring to their concept of (loop) quantum gravity: They start with loops from the outset and show that since spin network states $\langle S|$ span the loop state space, it follows that any ket state $|\psi\rangle$ is uniquely determined by the values of the S-functionals on it, namely of the form

$$\psi(S) := \langle S | \psi \rangle.$$

To be more precise, Rovelli and Smolin take embedded spin networks rather than the usual spin networks, i.e. they take the latter plus an immersion of its graph into a 3-manifold. Considering then, the equivalence classes of embedded oriented spin networks under diffeomorphisms, it can be shown that they are to be identified by the knotting properties of the embedded graph forming the network and by its colouring (which is the labelling of its links with positive integers). When generalizing this concept even further, a network design may be introduced as a conceptual approach towards pre-geometry based on the elementary concept of *distinctions*, as Louis Kauffman has shown.[4] In particular, space-time can be visualized as being produced directly from the operator algebra of a distinction. If thinking of distinctions in terms of 1-0 (or yes-no) decisions, we have a direct link here to information theory, which has been discussed recently again with a view to holography [47]. Ashtekar and Krasnov have noted this aspect already when deriving the celebrated Bekenstein-Hawking formula in applying loops to black holes. Referring to punctured horizons they can show that each set of punctures gives actually rise to a Hilbert space of (Chern-Simons) quantum states of the connection on the horizon. Be $P = \{j_{p1} ... j_{pn}\}$ the set of punctures, and $H_p$ the respective Hilbert space. Then dim $H_p \sim \Pi_{j_p \in P}(2j_p + 1)$, and the entropy of the black hole is simply given by $S_{bh} = \ln \sum_P \dim H_P$. The edges of spin networks can be visualized then as flux lines carrying area. With each given configuration of flux lines, there is a finite-dimensional Hilbert space describing the quantum states associated with curvature excitations initiated by the punctures of the horizon. Because the states that dominate the counting correspond to punctures all of which have labels $j = \frac{1}{2}$, each microstate can be thought of as providing a yes-no decision or an elementary „bit" of information. Hence, the reference to Wheeler's „It from Bit". (Paola Zizzi has tried to generalize this within the conception of inflationary cosmology, and terms this „It from Qubit".) Further details are given in [42]. See also [38], [39] for a recent review of the basic ideas as well as [1], [35].

---

[4] L.H.Kaufmann: Knots and Physics, 2nd ed., World Scientific, Singapore etc., 1993, 459sq.

Generalized spin networks can be used now for lattice gauge theory and for non-perturbative quantum gravity. In the former case, they turn out as products of Wilson loops. In the latter case, it can be found that the space of diffeomorphism invariant states is spanned by a basis which is in one-to-one correspondence with embeddings of spin networks. Remember that according to the standard terminology, a loop in some space $\Sigma$, say, is a continuous map $\gamma$ from the unit interval into $\Sigma$ such that $\gamma(0) = \gamma(1)$. The set of all such maps will be denoted by $\Omega\Sigma$, the loop space of $\Sigma$. Given a loop element $\gamma$, and a space $L$ of connections, we can define a complex function on $L \times \Omega\Sigma$, the socalled *Wilson loop* such that

$$T_A(\gamma) := (1/N) \, Tr_R \, P \exp \int_\gamma A.$$

Here, the path-ordered exponential of the connection $A \in L$, along the loop $\gamma$, is also known as the holonomy of A along $\gamma$. The holonomy measures the change undergone by an internal vector when parallel transported along $\gamma$. The trace is taken in the representation R of G (which is the Lie group of Yang-Mills theory), N being the dimensionality of this representation. The quantity measures therefore the curvature (or field strength) in a gauge-invariant way. (I am following here the terminology as given by Renate Loll in [2]. See also [8], [25].)

Over a given loop $\gamma$, the expectation value $<T(\gamma)>$ turns out to be equal to a knot invariant (the „Kauffman bracket") such that when applied to spin networks, the latter shows up as a deformation of Penrose's value $V(\Gamma)$. This is mainly due to the fact that

$$<T(\gamma)> = K^k(\gamma) = (1/Z) \int d\mu(A) \exp (...) T(\gamma, A).$$

So, for any spin network $\Gamma$ (replace $\gamma$ by $\Gamma$), the old relation holds up to regularization. Hence, spin networks are deformed into quantum spin networks (which are essentially given by a family of deformations of the original networks of Penrose labelled by some deformation parameter q).

There is also a simplicial aspect to this: Loop quantum gravity provides for a quantization of geometric entities such as area and volume. The main sequence of the spectrum of area e.g., shows up as $A = 8\pi\gamma\hbar G \sum_i (j_i(j_i + 1))^{1/2}$, where the j's are half-integers labelling the eigenvalues. This quantization shows that the states of the spin network basis are eigenvalues of some area and volume operators. We can say that a spin network carries quanta of area along its links, and quanta of volume at its nodes. A quantum space-time can be decomposed therefore, in a basis of states visualized as made up by quanta of volume which in turn are separated by quanta of area (at the intersections and on the links, respectively). Hence, we can visualize a spin network as sitting on the dual of a cellular decomposition of physical space.

The definition of a *spin foam* now is very much alike the one for a spin network, only one dimension higher. A spin foam is essentially taking one spin network into another, of the form F: $\Psi \to \Psi'$. Just as spin networks are designed to merge the concepts of quantum state and geometry of space, spin foams shall serve the merging of concepts of quantum history and geometry of space-time. Very much like Feynman diagrams do, also spin foams can be used to evaluate information about the history of a transition of which the amplitude is being determined. Hence, if $\Psi$ and $\Psi'$ are spin networks with underlying graphs $\gamma$ and $\gamma'$, then any spin foam F: $\Psi \to \Psi'$ determines an operator from $L^2(A_\gamma/G_\gamma)$ to $L^2(A_{\gamma'}/G_{\gamma'})$ denoted by o such that

$$\langle \Phi', o\, \Phi \rangle = \langle \Phi', \Psi' \rangle \langle \Psi, \Phi \rangle$$

for any states $\Phi, \Phi'$. The evolution operator Z(M) is a linear combination of these operators weighted by the amplitudes Z(o). Obviously, we can define a category with spin networks as objects and spin foams as morphisms. (We follow here the terminology of John Baez, for more details see [42] and also [2], [4].) Choosing BF theory as a simple case for the discussion, one can show (as Baez has actually done, cf. [42]) that there is a starting point for an explicit theory of emergence being developed on a line known from Topological Quantum Field Theory (TQFT). What we essentially do here is the following: Define space-time as a compact oriented cobordism M: S → S', where S, S' are compact oriented manifolds of dimension n-1. (Recall that two closed (n-1)-manifolds X and Y are said to be cobordant, if there is an n-manifold Z with boundary such that ∂Z is the disjoint union of X and Y.) Choose a triangulation of M such that the triangulations of S, S' with dual 1-skeletons $\gamma, \gamma'$ can be determined. The basis for gauge-invariant Hilbert spaces is given by the respective spin networks. Then the evolution operator Z(M): $L^2(A_{\gamma'}/G_{\gamma'}) \to L^2(A_\gamma/G_\gamma)$ determines transition amplitudes $\langle \Psi'$, Z(M) $\Psi \rangle$ with $\Psi, \Psi'$ being spin network states. Write the amplitude as a sum over spin foams from $\Psi$ to $\Psi'$: $\langle\ ,\ \rangle = \sum_{F:\Psi \to \Psi'} Z(F)$ plus composition rules such that Z(F') o Z(F) = Z(F' o F). This is a discrete version of a path integral. Hence, re-arrangement of spin numbers on the „combinatorial level" is equivalent to an evolution of states in terms of Hilbert spaces in the „quantum picture" and effectively changes the topology of space on the „cobordism level". This can be understood as a kind of *manifold morphogenesis* in time. These results can also be formulated in the language of category theory: As TQFT maps each manifold S representing space to a Hilbert space Z(S) and each cobordism M: S → S' representing space-time to an operator Z(M): Z(S) → Z(S') such that composition and identities are preserved, this means that TQFT is a functor Z: nCob → Hilb. The non-commutativity of operators in quantum theory corresponds to non-commutativity of composing cobordisms, and adjoint operation in quantum theory turning an operator A: H → H' into A*: H' → H corresponds to the operation of reversing the roles of past and future in a cobordism M: S $\to$ S' obtaining M*: S' $\to$ S.

Note that „time" shows up here not as a function, but as a manifold (although arrows are used). This is particularly interesting, because with a view to what Barbour tells us about the „absence" of time [5], this means that the concept of time is intrinsically included here as a *pragmatic ordering principle* for localizing topology changes. This is similar to what Prigogine calls the „age of a system", which is roughly a frequency of formations of new structures in a system making the latter more complex. Time as a convention then, would be an approximate „average" over such ages. Hence, time shows up as being associated to a kind of measuring device for local complexity gradients. So what we have in the end, is a rough (and simplified) outline of the foundations of emergence, in the sense that we can localize the fine structure of emergence (the re-arrangements of spin numbers in purely combinatorial terms being visualized as a motion-in-itself) and its results on the „macroscopic" scale (as a change of topology being visualized by physical observers as a motion-for-itself). But note also that space and time, in the classical sense, are obviously absent on the fundamental level of the theory, but they can be recovered as concepts when tracing the way „upward" to macroscopic structures. In other words: Even as a gross average feature for „shortsighted" human scientists (as Penrose indicates it at the end of his first twistor paper), space and time would nevertheless turn up as (philosophical) categories of concepts, simply, because the meaning of these categories is well-adapted to what humans actually perceive of their world (and communicate to other humans). This is in fact, a point, where Barbour's argument seems to break down (if discussed within this philosophical perspective): What he essentially shows in his book is that quantum theory, *in so far as it is foundational*, describes a part of what has been called non-being (or substance) in former times. Hence, there is neither space nor time in *real* terms ( = realiter, i.e. with respect to what there is in an absolute sense of the world's foundation), but there *is* space and time in *modal* terms ( = modaliter, i.e. with respect to what humans perceive of their world). The former refers to substance, or, alternatively, to what *speculative* philosophy is all about. The latter refers to the physical world, or, to what *sceptical* philosophy is all about. The one relies on theoretical speculation according to what we know - speculating about the foundation of the world, which is outside (logically „before") the world, and of which we are not a part therefore, and hence, about which we cannot actually know anything. The other refers to the empirical world, about which, with the help of experiments, we can obtain knowledge, in fact. Obviously, in terms of physics, the first (speculative) aspect is corresponding to physical theory, in so far as it is foundational. The second (sceptical) aspect corresponds to physical theory, in so far as it is empirical.

# 3  Quantum Computation

Recently, Bieberich has introduced a model for a non-local quantum evolution of entangled ensemble states in neural nets [6] which can also be used for spin nets, respectively. A system of qubits entangled in a pure state on a single node is considered evolving by unitary operation. The system decoheres and communicates its computational result to classical channels. Bieberich calls the unitary operation on the ensemble of qubits „mutation", if a coherent quantum state A evolves to another state A'. In this sense, mutation denotes a NOT-operation, and the respective quantum state the „genotype" (which remains pure if not measured). The phenotypic manifestation of this genotype proceeds by measurement of the system of entangled qubits due to environmental interactions. Measurement is actually being achieved by a translation of the form A $\rightarrow$ a, A' $\rightarrow$ a', where the translator is a set of operators which is composed of „readout" by measurement and „reset" of entanglement (the latter by a Hadamard operation or qubit rotation) [6], [11], [17], [22], [23], [37]. Bieberich shows in detail how such a network can filter and select the coupling of qubit states at the moment of translation by a choice of channel procedure.

The general idea is that some intrinsic dynamics trigger quantum oscillations which are essentially periodic fluctuations (or: vibrational modes) of the unitary operator between two pure states of entangled qubits with orthogonal eigenstates corresponding to those of the pointer states (a, a') within the environment. As this is basically addressed to the problem of consciousness, when applied to neuronal nets, this can help to describe resonance states of the network with respect to its environment such that conscious states including memory can be discussed in terms of the Hameroff-Penrose model [6], [47]. Because the fractal re-compression of communicated cbits onto the quantum register of single nodes and the average time delay between input and output (as measured from a macroscopic perspective) reproduce empirically relevant numerical values, the conclusion would be that the Universe itself, if visualized in terms of a spin network's vibrational modes, works with a tact sequence which is inherited by macroscopic (classical) structures. Hence, the expression „proto-consciousness" [47]. If using a Hopfield network matrix [6], it can be shown that a string of node couplings (knottings) is an elementary aspect of a resonance model for loop knots with one common knotting node for each single loop (which is referring here to loops in the sense that rows and columns of the Hopfield matrix are being closed by linked elements).

So far so good as to the immediate implications: Obviously, the spin network would show up as a system of channels which opens lines of communication (or information transport) in turn being processed as quantum computation on a fundamental level, very much in the sense of a process of directed percolation eventually resulting in macroscopic (classical) structures emerging from that quantum computation. In a way, one could think of a process which triggers a spontaneous onset of „gelation" (whose product is the classical world as we know it). This idea is not especially original by now, see e.g. [27], [20], [21], and also [15]. However, there remains a somewhat unsatisfactory point: the intrinsic role of time.

In fact, these „vibrational modes" of quantum computation on the fundamental level of physics should not actually refer to time in a classical sense. With that we mean a conventional time which is being created as suitable superposition of many multiples of the Planck time [47] (for another perspective towards the Planck level see also [46]). Because, as we have seen in the preceeding section, the conventional use of a concept of time is not consistent with respect to what we should expect of a fundamental level which is the foundation of our classical world so that „worldly" categories are actually absent from that level. However, this problem has not been addressed properly so far. On the contrary, the classical concept of time has entered the discussion from the beginning on, in a somewhat unnoticed manner.[5] Recently, the concept of time (and of causality) has become more controversial [5], [19], [27], [36].

The point is that transport problems like percolation which can in principle be modelled by means of a cellular automata architecture [10], [15], [16], [18], as well as „history" problems within the framework of variational principles, essentially emerge from a classical background, and have been re-introduced to quantum problems subsequently. But although speaking of „vibrational modes" and/or quantum „oscillations" refers to a language which has temporality implicit all the time (and it cannot be otherwise, because when thinking at all, humans are classical anyway, they can only perceive and express classical concepts therefore) – this is basically being visualized as a kind of intrinsic and *non-local* (non-spatial) *as well as a-temporal* motion. There is not even any necessity as to a partial ordering on the category of fundamental states. Basically one should actually start from a completely disordered set of some kind from which an ordered set is eventually emerging. The technical question is as to the appropriate choice of an adequate functor achieving this [45], [49].

This problem can be clarified somewhat when thinking of the path integral approach to „histories" which is also reminiscent of the Wilson loop terminology mentioned above: The idea as it has been introduced by Feynman has been to argue that the probability for a system's transition from a state A to another state

---

[5] Not completely unnoticed in fact, but in a series of early papers by Heisenberg, Born, and Jordan e.g., when discussing unitary operations and also variational principles, no such problem is being mentioned. See K.B.Beaton and H.C.Bolton: A German Source-book in Physics, Clarendon Press, Oxford, 1969, 138-162. In the celebrated textbook of Dirac's, it is explicitly stated that „ (the) variation with time of the Heisenberg dynamical variables may ... be looked upon as the continuous unfolding of a unitary transformation." But the t-variable is not really questioned in the first place when motivating the introduction of the dynamics. (P.A.M. Dirac: The Principles of Quantum Mechanics ($4^{th}$ ed.), Clarendon Press, Oxford, 1958, 129.) Even Feynman, in his famous lecture on variational principles, makes no difference when introducing quantum systems. (The Principle of Least Action, ch.19, in: R.P.Feynman, R.B.Leighton, M.Sands, The Feynman Lectures on Physics, vol. 2, CIT, 1963; German edition, Muenchen, 1987.)

B can be expressed as the absolute value squared of the respective path integral summing over all possible connections from A to B, basically of the form $\int \exp(iS/\hbar)$, where S is the appropriate action. However, this action S itself is given by another integral, of the form $\int L\, dt$. It is argued then that „on the quantum level", all possible transitions from A to B take place *at the same time*, while in classical terms one path is being selected (the „classical path" which is actually being observed then) owed to interference effects which cancel out all other paths not satisfying the variational condition $\delta S = 0$. This may be true above the Planck level, but the question of what t actually means *at this level* is still an open topic. Consequently, carrying this aspect over to actions showing up in other forms of path integrals (such as the Wilson loop integrals) – save perhaps the Chern-Simons action which is constituting a TQFT rather - it is apparent that those cannot be entities effective on the fundamental level of physics. In fact, when discussing loops and knots, Pullin [29] founds the existence of constraints on the fact that lapse and shift do not enter with a time derivative so that their conjugate momenta are zero. (A similar argument is taken up by Barbour [5] in his book on the absence of time where he stresses the significance of the Hamiltonian constraint.) After introducing Ashtekar (connection) variables, Pullin demonstrates how Wilson loops show up as an infinite parameter family of gauge invariant functionals of a connection. They are not a solution of the diffeomorphism constraint, but if formed with an Ashtekar connection, they do solve the Hamiltonian constraint. In particular, the Kauffman bracket shows up then as a solution of the latter with a cosmological constant. Wilson loops constructed with smooth loops turn out to be eigenstates of the metric operator. Hence, if also thinking of the various connotations of Wilson loops in a very general context of physical fields (e.g. in biophysics, as Louis Kauffman has discussed in his book on knots, see also more recently Patel on DNA structure [28]), it is straightforward to expect their relevance on the first level above the Planck level (thus on the „classical side" of physics rather than on a fundamental level). They actually become also relevant for Smolin's concept of cosmological selection [41], [42].

There is a final point with respect to spin foams: Usually, they are being introduced as formal time development of spin networks, replacing the trivalent graphs by a succession of them, visualized in terms of appropriate 2-complexes. Practically, as we have seen in the preceeding section, a path integral analogue is being introduced for defining the amplitude as summing over spin foams so as to give the explicit unfolding of an evolution of spin nets. Hence, the impression is implied that „in between" nets „time is running" somehow. Sometimes, even light cones are being attached (on a level of physics where there should be neither light in the strict sense, nor space and time altogether) [27]. A possible way out could be to visualize spin foams as expression for the above mentioned „vibrational modes" constituting the intrinsic motion as performed in terms of quantum computation. The detailed „mode structure" could be thought of then as being inheri-ted by the macroscopic level so as to display the actual time (being observed on the latter) as immanent property of the respective cobordism model. Consequently, time (as well as space) would show up as necessary categories of human thinking on the classical level of which humans are a product in the first place. Time would be nothing than an ordering parameter then having the task to coordinate digital thinking (cf. [5], [42], [43]). As I have discussed in more detail in [42], one hope may be to utilize recent insight into (mathematical) category and topos theory in order to find a sound onto-epistemic basis to treat these problems. See also [9], [12] – [14], [25], [26]. More aspects of spin foams will be discussed in the forthcoming [49]. There is also an interesting recent development concerning the relationship between spin foams and Feynman diagrams [30], [31], which might enhance a re-interpretation of the (formal) time axis in the latter diagrams. As a base text for this may serve the very nice and comprehensive treatment by Veltman [40]. For a recent review see also [32]. The promising n-categories approach by Baez and Dolan [3] is very much on this line of thinking.

## 4   Conclusions

The main difficulty to start physics from a completely non-spatial, non-temporal, and thus non-causal structure underlying it on a fundamental level is, as we have seen by now, the specific structure of our thinking itself. This is relying on a progressive ordering of everything what is being observed, in a sequential arrangement of digitalized moments of thought. This is nothing but a question of the actual perceptive (and cognitive) capacity of humans which in turn is an outcome of biological evolution on this very planet. An old idea of speculative philosophy was to develop models for expressing something about which we cannot collect knowledge, because it is falling out of the framework according to which our knowledge is being structured in the first place. Because many fields of physics are dealing with things which fall out of the framework of our macroscopic sense perceptions, it was a common idea in physics that this kind of abstraction would have been well established in practice by now. But, as it turns out, this scheme is likely to break down when trying to assess the truly fundamental level of physics. First of all, this introduces a split into physics manifesting an explicit boundary of knowledge in principle. Hence, theories dealing with the conceptual modelling of that fundamental level are practically detached (in principle) from their empirical reference.[6] Second, there is a fundamental deficiency of language itself which is creating explicit difficulties to actually talk about structures and phenomena which are non-spatial and non-temporal. Probably, it is not a coincidence that all these problems converge with a recent renaissance of consciousness research, because it is very likely that some time there will be a theory which is *onto-epistemic* in the first place such that the modelling itself (the thinking about things) is appropriately adapted to what is being thought about from the beginning on. After all, human thinking (together with humans) is nothing but a product of the natural processes which it is setting out to model.

---

[6] Note, as an example, that it is this reason why Fay Dowker cannot really claim from Julian Barbour that he should give hints as to innovative experiments confirming his approach. Stating the absence of time on a fundamental level is withdrawing the discussion from any empirical claims, because it is a statement of foundational rather than empirical physics, and therefore it is not testable in principle. Very much like speculative philosophy as related to sceptical philosophy, foundational physics has only to meet the condition of consistency with what is already known from empirical physics. Hence, speculating does not mean to invent arbitrary things, this is instead the (nevertheless reasonable) task of the various arts.

Until now, physical theories seem to come in two classes, in one class of theories which are ontologically correct, and in one class of theories which are epistemologically correct. An example for the former would be loop theory, because it is quite obvious why an ontologically adequate theory should be one without any reference to a background space. An example for the latter would be superstrings and/or M theory, because obviously, it is adequate to describe the world (as it is attainable to human empirical research) by means of dimensions of some kind. (Note that if space and time are absent on a fundamental level, then so are all other dimensions.) A promising candidate for a truly onto-epistemic theory would be non-perturbative M theory [34]. Probably, the various controversial contributions to an adequate conceptualization of the fundamental level of physics will be solved not before some such theory is actually being achieved.

## Acknowledgements

This paper owes a lot to the cooperation with Paola Zizzi. I thank her for stimulating and illuminating discussions. Some further results of this cooperation will be published in two forthcoming papers [45], [49].